\documentclass{nature}
\usepackage{epsfig}
\usepackage{amsmath}
\usepackage{amssymb}

\DeclareMathOperator*{\Tr}{Tr}

\DeclareMathOperator*{\diag}{diag}
\DeclareMathOperator*{\arctanh}{arctanh}

\begin{document}

\title{Topological Quantum Phase Transition in Synthetic Non-Abelian Gauge Potential: Gauge Invariance and Experimental Detections}
\author{Fadi Sun$^{1,2\dagger\star}$, Xiao-Lu Yu$^{1\dagger}$, Jinwu Ye$^{2,3\dagger}$,
Heng Fan$^{1\dagger}$, and Wu-Ming Liu$^{1\dagger}$}

\maketitle

\begin{affiliations}
\item
Beijing National Laboratory for Condensed Matter Physics,
Institute of Physics, Chinese Academy of Sciences,
Beijing 100190, China

\item
Department of Physics and Astronomy,
Mississippi State University, P. O. Box 5167,
Mississippi State, MS 39762, USA

\item 
Department of Physics, Capital Normal University,
Beijing 100048, China

$^\dagger$These authors contributed equally to this work.

$^\star$e-mail: sfd@iphy.ac.cn

\end{affiliations}

\begin{abstract}
The method of synthetic gauge potentials opens up a new avenue for our understanding and discovering novel quantum states of matter.
We investigate the topological quantum phase transition of Fermi gases trapped in a honeycomb lattice in the presence of a synthetic non-Abelian gauge potential.
We develop a systematic fermionic effective field theory to describe a topological
quantum phase transition tuned by the non-Abelian gauge potential and explore its various important experimental consequences.
Numerical calculations on lattice scales are performed to compare with the
results achieved by the fermionic effective field theory.
Several possible experimental detection methods of topological quantum phase transition are proposed.
In contrast to condensed matter experiments where only gauge invariant quantities can be measured,
both gauge invariant and non-gauge invariant quantities can be measured by experimentally generating various non-Abelian gauges
corresponding to the same set of  Wilson loops.
\end{abstract}

A wide range of atomic physics and quantum optics technology
provide unprecedented manipulation of a variety of intriguing quantum phenomena.
Recently, based on the Berry phase \cite{Berry1984}
and its non-Abelian generalization \cite{Wilczek1984},
Spielman's group in NIST has successfully generated
a synthetic external Abelian or non-Abelian gauge potential coupled to neutral atoms.
The realization of non-Abelian gauge potentials in quantum gases opened a new avenue in cold atom physics
\cite{Lin1,Lin2,Lin3,Lin4,Sau2011,Anderson2012,Niu2010,Dalibard2011}.
It may be used to simulate various kinds of relativistic quantum field theories \cite{Bermudez2010L,Mazza2012},
topological insulators \cite{Kane2010,Zhang2011}, graphene \cite{Neto2009,Vozmediano2010},
and it may also provide new experimental systems in finding Majorana fermions \cite{HPu2012,Melo2012}.

Recently, there have been some experimental \cite{Soltan-Panahi2011,Tarruell2012} and theoretical activities \cite{Zhu2007,Jiang2006,Ye2008NPB,tqcp} in
manipulating and controlling of ultracold atoms in a honeycomb optical lattice.
Bermudez {\it et al.} \cite{Bermudez2010} studied Fermi gases trapped in a honeycomb optical lattice in the presence of a synthetic $SU(2)$ gauge potential.
They discovered that as one tunes the parameters of the non-Abelian gauge potential,
the system undergoes a topological quantum phase transition (TQPT) from the $ N_D=8 $ massless Dirac zero modes phase to a $ N_D=4$ phase.
However, despite this qualitative picture, there remain many important open problems. 
In this work, we address these important problems.
We first determine the phase boundary in the two parameters of the non-Abelian gauge potential,
and provide a physical picture to classify the two different topological phases and the TQPT from the magnetic space
group (MSG) symmetry \cite{Wen2004,Jiang2006,Ye2008NPB}. Then we develop a systematic fermionic effective field theory (EFT) to describe such a TQPT
and explore its various important experimental consequences.
We obtain the critical exponents at zero temperature which are contrasted with a direct numerical calculation on a lattice scale.
We derive the scaling functions for the single particle Green function, density of states, dynamic compressibility,
uniform compressibility, specific heat and Wilson ratio. A weak short-ranged atom-atom interaction is irrelevant near the TQPT, but the disorders
in generating the non-Abelian gauge fields are relevant near the TQPT.
We especially distinguish gauge invariant physical quantities from non-gauge invariant ones.
When discussing various potential experimental detections of the topological quantum phase transition,
we explore the possibilities to choose different gauges to measure both gauge invariant and non-gauge invariant physical quantities.
We stress the crucial differences between the TQPT discussed in this work and some previously known TQPTs.

\section*{Results}

\subsection{Wilson loops, topological phases and phase boundary}
The tight-binding Hamiltonian for Fermi gases trapped in a two-dimensional (2D) honeycomb optical lattice in the presence of a non-Abelian gauge potential
[Fig.~1(a)] is:
\begin{eqnarray}
\mathcal{H}_0
\!=\!
-t\!\sum\limits_{\langle i,j\rangle}
c_A^\dagger(i\sigma)
U_{ij}^{\sigma\sigma'}
c_B(j\sigma')+h.c.,
\label{eq:TBU}
\end{eqnarray}
where $t$ is the hopping amplitude,
$\langle i,j\rangle$ means the nearest neighbors,
and $c_{A}^\dagger(i\sigma)$,$c_{B}^\dagger(i\sigma)$
($c_{A}(i\sigma)$,$c_{B}(i\sigma)$) create (annihilate) a fermion
at site ${\bf r}_i$ of $A$- and $B$-sublattice with spin $\sigma$.
The unitary operator $ U $ is related to the non-Abelian gauge potentials ${\bf A}$
by the Schwinger line integral along the hopping path $\mathcal{P}\exp(i\frac{e}{h}\int{\bf A}\cdot d{\bf l})$.
For simplicity, we choose a lattice translational invariant gauge [Fig.~1(a)] where
the $U_{ij}=U_{i-j}\equiv U_\delta, \delta=1,2,3 $. In the momentum space, the Eq.~\eqref{eq:TBU}
can be written \cite{Bermudez2010} as $ H= \sum_{\vec{k}} \Psi^{\dagger}_{a} ( \vec{k} ) H_{ab}(\vec{k}) \Psi_{b}( \vec{k} ) $
where $ a, b $ stands for the two sublattices $ A, B $ and also the the two spin indices $ \sigma  $.
In the specific gauge in Fig.~1(a), $U_1=e^{i\alpha\sigma_x}$, $ U_2=1 $ and $U_3= e^{i\beta\sigma_y} $,  where the
$\sigma_x$ and $\sigma_y$ are Pauli matrices in the spin-components 
(The more general case $ U_1=e^{i\alpha\sigma_x}, U_2=e^{i\gamma \sigma_z}, U_3=e^{i\beta\sigma_y} $ can be similarly discussed).
The gauge invariant Wilson loop \cite{Bermudez2010}  around an elementary hexagon is $ W(\alpha,\beta)=2-4\sin^2\alpha\sin^2\beta $
which stands for the non-Abelian flux through the hexagon. However, in contrast to the Abelian gauge case on a lattice \cite{Jiang2006},
the $ W(\alpha,\beta) $ is not enough to characterize the gauge invariant properties of the system.
One need one of the three Wilson loops $ W_{1,2,3} $ around the 3 orientations of two adjacent hexagons to achieve the goal.
In the gauge in Fig.~1(a), $ W_{1}(\alpha,\beta)=2-4\sin^2 2\alpha \sin^2\beta$,
$W_{2}(\alpha,\beta)=2-\sin^2 2\alpha \sin^2 2\beta$, and $W_{3}(\alpha,\beta)=2-4\sin^2 \alpha \sin^2 2\beta  $.
The gauge invariant phase boundary in terms of $ W $ and $ W_1 $ is shown in Fig. 4(a). 
The $ W=\pm 2 $ and $ | W | < 2 $ correspond to Abelian regimes and non-Abelian regimes respectively.
Only in the Abelian case $ W_{1} = W_2 = W_3= 2 $.
The fact that $ W_{1} \neq W_2 $ or $ W_{1} \neq W_3  $ in the non-abelian case shows that
the $ 2 \pi/3 $ rotation symmetry around a lattice point 
(or $ \pi/3 $ symmetry around the center of the hexagon)
is generally broken, even the translational symmetry is preserved by the non-Abelian gauge field.
It is easy to see that the $\alpha$ is gauge equivalent to $\pi \pm \alpha$
and $\beta$ is gauge equivalent to $\pi \pm \beta$, so we can restrict $\alpha$ and $\beta$ in the region $[0,\pi]$.
It is important to stress that in principle, the cold atom experiments \cite{Lin1,Lin2,Lin3,Lin4,Sau2011,Anderson2012,Niu2010,Dalibard2011}
can generate various gauges corresponding to the same $ W $ and $ W_{1,2,3} $, so the gauge parameters $\alpha$ and $\beta$ are experimentally adjustable. 
This fact will be important in discussing experimental detections of the TQPT.

The spectrum of $\mathcal{H}$ consists of four bands given by
\begin{eqnarray}
\epsilon_{1\pm}({\bf k})\!=\!\pm t\sqrt{b_{\bf k}\!-\!2\sqrt{d_{\bf k}}},~
\epsilon_{2\pm}({\bf k})\!=\!\pm t\sqrt{b_{\bf k}\!+\!2\sqrt{d_{\bf k}}}
\label{eq:E},
\end{eqnarray}
where the $ b_{\bf k} =3+2\cos\alpha\cos{(k_{1}a)} +2\cos\beta\cos{[(k_{1}+k_{2})a]} +2\cos\alpha\cos\beta\cos{(k_{2}a)}$ and
$ d_{\bf k}=(1-\cos^2\alpha\cos^2\beta)\sin^2{(k_{2}a)}+\sin^2\alpha\sin^2{(k_{1}a)}
-2\sin^2\alpha\cos\beta\sin{(k_{1}a)}\sin{(k_{2}a)}+\sin^2\beta\sin^2{[(k_{1}\!+\!k_{2})a]} +2\cos\alpha\sin^2\beta\sin{[(k_{1}\!+\!k_{2})a]}\sin{k_{2}a}$
with $k_{1}=3k_{x}/2-\sqrt{3}k_{y}/2 $ and $k_{2}=\sqrt{3}k_y$.
In the following, we focus on the most interesting half-filling case. At the half filling,
the spectrum is particle-hole symmetric,
the $\epsilon_{1\pm}$ ($\epsilon_{2\pm}$) describe the two low  (high) energy bands.
By solving $\epsilon_{1-}({\bf k})=0$ for ${\bf k}$ which can be expressed as the roots of a quartic  equation,
we obtain all the zero modes in analytic forms. For simplicity, we only show the number of the zero modes $ N_D $ in Fig.~1(b) for general $ \alpha, \beta $.
Especially, the phase boundary in Fig.~1(b) separating $ N_D=8 $ from the $ N_D=4 $ zero modes  is determined by setting
the discriminant of the quartic equation to be zero.
As the gauge parameters $\alpha$ and $\beta$ change from $0$ to $\pi$, the system undergoes a topological quantum phase transition (TQPT)
from the $ N_D=8 $ massless Dirac zero modes phase in the yellow regime to a $ N_D=4$ phase in the green regime shown in the Fig.~1(b).
Along the dashed line in the Fig.~1(b), the TQPT at $ ( \alpha=\pi/2, \beta_c=\pi/3 ) $ is induced by changes in Fermi-surface topologies  shown in the Fig.~2(a)-(d).

\subsection{Classification of the topological quantum phase transition by the magnetic space group}
  Time-reversal symmetry indicates that the only two Abelian points are $ W = \pm 2 $
  which correspond to no flux and the $ \pi $ flux respectively.
  For a Abelian flux $ \phi=1/q $, the MSG dictates there are at least $ q $ minima in the energy bands \cite{Jiang2006}.
  If there exists Dirac points (zero modes), the MSG dictates there are at least $ q $ Dirac points in the energy bands.
  All the low energy modes near the $ q $ Dirac points construct a $ q $ dimensional representation the of MSG.
  Due to the time-reversal symmetry, the Dirac points always appear in pairs.  When counting the two spin components,
  each Dirac zero mode was doubly degenerate, so they are counted as $ N_D=4q $ Dirac points.
$W=-2$ corresponds to the $ q=2 $ case where there are  $ N_D=8 $ Dirac zero modes.
It is the $\pi$ flux Abelian point locating at the center in the Fig.~1(b).
$W=2$ case corresponds to the $ q=1 $ case where there are $ N_D=4 $ Dirac zero modes.
It is just the graphene case \cite{Neto2009} running along the 4 edges of the square in the Fig.~1(b) ( See Sec. 8 ).
Obviously, the $ W= \pm 2 $ have different Fermi surface topologies, so there must be a TQPT  separating the two extremes.
It is the non-Abelian gauge field which tunes between the two
Abelian points landing in the two different topological phases, induces the changes in Fermi-surface topologies and drives the TQPT.
The different phases  across the TQPT are  characterized by different topologies of Fermi surface \cite{Volovik2003} instead of being classified by different symmetries, so they are beyond Landau's paradigm.

\subsection{The low energy effective field theory}
Based on the physical pictures shown in the Fig.~2,
we will derive the effective action near the TQCP
at $ ( \alpha=\pi/2, \beta = \pi/3 ) $ by the following procedures:
(1) Perform an expansion around the critical point $ \beta= \beta_c= \pi/3 $
and the merging point $ {\bf P}=(\frac{\pi}{2},-\frac{\pi}{2\sqrt{3}}) $
(or equivalently  ${\bf Q}=-{\bf P} $):
$H(k_x,k_y,\Delta)=H^{0}_{P}+H^{x}_{P}q_x+H^{y}_{P}q_y
+\frac{1}{2}H^{xx}_{P}q^{2}_x+H^{\Delta}_{P}\Delta+\cdots$
where the ${\bf k= P +q},~ |{\bf q}|\ll 1/a $ and the $ \Delta\propto\beta_c -\beta,~ |\Delta| \ll \beta_c $.
(2) Perform a counter-clockwise rotation $ R_{\pi/6} $ by $\pi/6$ around the point $ {\bf P} $ to align the $ q_x $ along the colliding direction.
(3) Diagonalize the $ H^{0}_{P} $ by the unitary matrix $ S_{P} $: $ S^{\dagger}_{P} H^{0}_{P} S_{P}=\diag(-2t,2t,0,0) $.
(4) Separate Hamiltonian into $2\times2$ blocks
in terms of high (low)  energy component $ \phi_H $ ($ \phi_L $), then
 adiabatically eliminate the high-energy bands around $ -2t $ and $ 2t $
 to obtain the effective low-energy two bands Hamiltonian around ${\bf q}=0$ acting on the low energy component $ \phi_L $.
 Finally, we obtain the effective Hamiltonian density in term of effective field $ \phi_L $ (See Method section)
\begin{eqnarray}
\mathcal{H}_{\rm{eff}}({\bf q})
= \phi^{\dagger}_{L}({\bf q}) [ v\hbar q_y\sigma_x+(\frac{\hbar^2q_x^2}{2m}+\Delta)\sigma_y ]  \phi_{L}({\bf q}),
\label{eq:aHeff}
\end{eqnarray}
where $v=\frac{3ta}{2\hbar}, m=\frac{\hbar^2}{3ta^2}, \Delta=\frac{\sqrt{3}t}{2}(\frac{\pi}{3}-\beta)$ and
the effective field $\phi_L=[\psi_1,\psi_2]^T$ is related to the original lattice fields by Eq.~\eqref{latticefields}.
When $\Delta<0$ ($\Delta>0$), it is in the $N_D=8$ ($N_D=4$) phase. We obtain the energy spectrum
$\epsilon_{\pm}({\bf q})=\pm\sqrt{(\hbar^2q_x^2/2m+\Delta)^2+v^2\hbar^2q_y^2}$
is quartic (diffusive) in the colliding direction ($q_x$ direction),
but linear (ballistic) in the perpendicular direction ($q_y$ direction) 
(Similar anisotropous were observed in the collision between two $ U(1) $ gauge vortices of two opposite
winding numbers $ \mu=\pm 1 $ in an expanding universe, see \cite{universe}).
Note that the Eq.~\eqref{eq:aHeff} and Eq.~\eqref{latticefields} were derived at a fixed gauge, namely along the dashed line in Fig.~1(b),
so the position of the merging point $ {\bf P} $  (or ${\bf Q}=-{\bf P} $)
will change under a gauge transformation.
This fact will play very important roles in the experimental detections of the TQPT
and will be discussed in details in the last section.

Applying the same procedures to the four ${\bf K}_{1,2,3,4}$ points in the Fig.~2,
we obtain the usual Dirac-type Hamiltonian for these points.
These four Dirac points stay non-critical through the TQPT,
so they just contribute to a smooth background across the TQPT.
In the following, we subtract the trivial contributions from the $ N_D=4 $ ``spectator''  fermions from all the physical quantities.
All the physical quantities should be multiplied by a factor of 2 to take into account
the two merging points $\bf P$ and ${\bf Q}=-{\bf P}$.

\subsection{The zero temperature critical exponents}
Now we investigate if there are any singular behaviors of the ground state energy across the TQPT.
So we calculate the gauge invariant ground state energy density
$\mathcal{E}(\Delta) = \tfrac{1}{4\pi^2}\int d^2{\bf q}\epsilon_{-}({\bf q}) $ and extract its non-analytic part.
Its 2nd derivative with respect to $\Delta$ around the critical point is  (See Method section)
\begin{align}
\mathcal{E}''(\Delta)\sim
\begin{cases}
-\frac{1}{\pi}\frac{\sqrt{2m}}{\hbar^2v}\sqrt{\Delta}, &{\rm for}~ \Delta>0, \\
\frac{\sqrt{2}}{3\pi^2}K(\frac{1}{2})\frac{\sqrt{2m}}{\hbar^2v}\frac{\Delta}{\sqrt{\Lambda}}, &{\rm for}~ \Delta<0,
\end{cases}
\label{cusp}
\end{align}
where $K(1/2)\approx 1.85$, $K(z)$ is the complete elliptic integral of the first kind,
and $ \Lambda$ is an ultraviolet energy cutoff in the integral.
We define $\mathcal{E}''(\Delta)-\mathcal{E}''(\Delta\!=\!0)\propto|\Delta|^{-\nu}$,
where $\nu$ is the critical exponent characterizing the TQPT.
Eq.~\eqref{cusp} shows that the $\mathcal{E}''(\beta)$ exhibits a cusplike behavior
with  the critical exponent $\nu_{+}=-1/2$ and  $\nu_{-}=-1$.
Obviously $ \mathcal{E}'''(\Delta) $ diverges like $ \Delta^{-1/2} $ as the $ \Delta \rightarrow 0^{+} $,
but approaches a constant as the $ \Delta \rightarrow 0^{-} $.
The TQPT is 3rd order continuous quantum phase transition.
In contrast, most conventional continuous quantum phase transitions are 2nd order.

We numerically calculate the ground-state energy density on the lattice scale
$\mathcal{E}_{\rm latt}(\beta)\!=\!\frac{1}{4\pi^2}\int_{\rm{BZ}}d^2{\bf k}\left[\epsilon_{1-}({\bf k})+\epsilon_{2-}({\bf k})\right] $
and illustrate its 1st, 2nd and 3rd derivatives in Fig.~3.
Indeed, the $\mathcal{E}''(\beta)$ exhibits a cusplike behavior near $\beta_c=\pi/3$ in Fig.~3(c).
Numerically, we obtain the critical exponent $\nu_{+}\!=\!-1.0$ and  $\nu_{-}\!=\!-0.5$
consistent with our analytical results Eq.~\eqref{cusp} (Note that $\Delta\propto \beta_c-\beta$).
This fact confirms that the effective Hamiltonian Eq.~\eqref{eq:aHeff} indeed captures the low energy fluctuations across the TQPT.

\subsection{Scaling functions at finite temperature}
At finite temperature $T$, the free energy density $\mathcal{F}$ is
\begin{equation}
\mathcal{F}
=-2k_BT\int\frac{d^2{\bf q}}{(2\pi)^2}\ln(1+e^{-\epsilon_{+}/k_BT})+\mathcal{E}(\Delta),
\label{eq:f}
\end{equation}
where the  $\mathcal{E}(\Delta)$ is the ground state energy density whose singular behaviors were extracted above.
It turns out all the singular behaviors are encoded in $\mathcal{E}(\Delta)$.
There is no more singularities at any finite temperature,
so the TQPT becomes a crossover at any finite $T$.
Following Ref.~\cite{Chubukov1994,annalscaling}, we can sketch the finite temperature phase diagram in the Fig. 4(b).
We can write down the scaling forms of the retarded single particle Green function,
the dynamical compressibility and the specific heat
\begin{align}
G^{R}(q_x,q_y,\omega)
&=\tfrac{\hbar}{k_B T}A_{i}
(\tfrac{\hbar\omega}{k_BT},
 \tfrac{\hbar q_x}{\sqrt{2mk_B T} },
 \tfrac{\hbar v q_y}{k_B T},
 \tfrac{|\Delta|}{k_B T}), \nonumber\\
\kappa^{R}(q_x,q_y,\omega)
&=\tfrac{\sqrt{2mk_B T}}{\hbar v}\Phi_{i}
(\tfrac{\hbar\omega}{k_B T},
 \tfrac{\hbar q_x}{\sqrt{2mk_B T}},
 \tfrac{\hbar v q_y}{k_B T},
 \tfrac{|\Delta|}{k_B T}),\nonumber\\
C_{v}
&=\frac{k_B\sqrt{2m}}{\hbar^{2}v}(k_B T)^{3/2}
\Psi_{i}(\tfrac{|\Delta|}{k_B T}),
\label{scaling}
\end{align}
where the subscript $i=1$ ($i=2$) stands for the $N_D=4$ ($N_D=8$) phase.
Note the anisotropic scalings in $ q_x $ and $ q_y $.

Although  the single fermion Green function $G^{R}$ is gauge dependent,
the single particle DOS $\rho(\omega)=-2\Tr\int d{\bf q} \Im G^{R}({\bf q},\omega)|_{T=0}
=\frac{\sqrt{2m\hbar\omega}}{\hbar v} \tilde{A}_i(\frac{\hbar\omega}{|\Delta|})$
is gauge-invariant.
The dynamic compressibility  and the specific heat are gauge invariant.
The uniform compressibility is given by
$\kappa_{u}=\kappa^{R} ({\bf q}\rightarrow0,\omega=0)
=\frac{\sqrt{2mk_BT}}{\hbar v} \Omega_i\left(\frac{|\Delta|}{k_BT}\right) $.
We have achieved the analytic expressions for  $\kappa_{u}(T)$ and $C_{v}$ (See method section).
Here, we only list their values in the three regimes shown in the Fig. 4(b).
For the uniform compressibility, we have
\begin{align}
\kappa_{u}\!=\!
\begin{cases}
0.22 \!\times\!
\frac{\sqrt{2m}}{\hbar^2 v}\frac{k_BT}{\sqrt{-\Delta}},
&{\rm for}~\Delta\ll-k_BT,\\
0.14\!\times\!
\frac{\sqrt{2mk_BT}}{\hbar^2 v},
&{\rm for}~|\Delta|\ll k_BT,\\
0.22\!\times\!
\frac{\sqrt{2m\Delta}}{\hbar^2 v}e^{-\frac{\Delta}{k_BT}},
&{\rm for}~\Delta\gg k_BT.
\end{cases}
\label{comp}
\end{align}
For the specific heat, we have
\begin{align}
C_V\!\!=\!\!
\begin{cases}
1.72\!\times\!\!
\frac{k_B\sqrt{2m}}{\hbar^2v}\frac{(k_BT)^{2}}{\sqrt{-\Delta}},
&{\rm for}~\Delta\ll-k_BT,\\
0.76\!\times\!
\frac{k_B\sqrt{2m}}{\hbar^2v}(k_BT)^{3/2},
&{\rm for}~|\Delta|\ll k_BT,\\
0.22\!\times\!
\frac{k_B\sqrt{2m}}{\hbar^2v}
\frac{\Delta^{5/2}}{k_BT}e^{-\frac{\Delta}{k_BT}},
&{\rm for}~\Delta\gg k_BT.
\end{cases}
\label{spec}
\end{align}

From Eq.~\eqref{comp} and Eq.~\eqref{spec},
we can form the Wilson ratio between the compressibility and the specific heat
$R_W(\frac{|\Delta|}{k_BT})
=k_B^2T\kappa_{u}/C_v
=\Omega_i(\frac{|\Delta|}{k_BT})/\Psi_i(\frac{|\Delta|}{k_BT})$
whose values in the three regimes in the Fig. 4(b) are
\begin{align}
R_W
=
\begin{cases}
0.12,
&{\rm for}~\Delta\ll-k_BT,\\
0.18,
&{\rm for}~|\Delta|\ll k_BT,\\
(\frac{k_BT}{\Delta})^2,
&{\rm for}~\Delta\gg k_BT.
\label{wilson}
\end{cases}
\end{align}

\subsection{Effects of interactions and disorders}
Now we consider the effects of a weak Hubbard-like short-range interactions
$ U \sum_{i\in A,B}n_{i\uparrow}n_{i\downarrow}$
on the TQPT in Fig. 4(b).
Following the standard renormalization group (RG) procedures in \cite{coul1,coul2} (See Method section),
we find the scaling dimension of $ U $ is $-1/2<0$,
so it is irrelevant near the TQPT at $ {\bf P} = - {\bf Q}$.
It was known \cite{coul1,coul2} that the $ U $, with the scaling dimension $ -1 < 0 $, is also irrelevant
near the Dirac  points at ${\bf K}_{1,2,3,4}$. So all the leading scaling behaviors will not be changed by the weak short-range interaction.
For the quenched disorders $\Delta_g$ in the gauge parameters $\alpha,\beta$,
following the RG procedures in \cite{coul1,coul2},
we find its scaling dimension is $1/2>0$,
so they are relevant to the TQPT at $ {\bf P} = - {\bf Q}$.
It was known \cite{coul1,coul2} that the $ \Delta_g $, with the scaling dimension $  0 $, is marginal
near the Dirac  points at ${\bf K}_{1,2,3,4}$.
This put some constraints on the stabilities of the laser beams generating the synthetic gauge field.
It would be interesting to look at the interplays between the strong repulsive or negative  $ U $ and
the non-Abelian gauge potentials near the TQPT.

\subsection{Gauge invariance and gauge choices in Experimental detections of the topological quantum phase transition}
Due to absence of symmetry breaking across a TQPT,
it remains experimentally challenging to detect a TQPT.
Very recently, the Esslinger's group in ETH \cite{Tarruell2012} has manipulated two time-reversal related Dirac points \cite{tqcp}
in the band structure of the ultracold Fermi gas of $^{40}$K atoms by tuning the hopping anisotropies
in a honeycomb optical lattice and identified the two Dirac zero modes via the momentum resolved interband transitions (MRIT).
As to be stressed in the disscussion section, in the present synthetic gauge potential problem,
the positions of the Dirac points and the two merging points $ {\bf P =- Q} $
shown in Fig.~2 are gauge-dependent,
so can be shifted by a gauge transformation. We expect that by tuning the orientations and intensity profiles of
the incident laser beams, various gauges corresponding to the same Wilson loops $ W $ and $ W_{1,2,3} $ can be
experimentally generated. So the MRIT measurement can still be used to detect the positions of the two merging points, the Dirac points and the TQPT at a fixed gauge. Then it can be repeatedly performed at various other experimentally chosen gauges to monitor the changes of these positions
as the gauge changes. However, the number of Dirac points $ N_D$ in the two different topological phases and the density of states $\rho(\omega)$ are gauge invariant. In principle, the number of Dirac points $N_D$ can be measured by Hall conductivities.
The $\rho(\omega)$ can be measured by the modified RF-spectroscopy \cite{rf1,rf2}.
There are previous experimental measurements on the specific heat of a strongly interacting Fermi gas \cite{heatcapacity}.
Very recently, Ku {\it et al.} \cite{compres} observed the superfluid phase transition in a strongly interacting $^6$Li Fermi gas by
presenting precise measurements of the compressibility $\kappa_{u}$  and the heat capacity $C_v$.
It was demonstrated that the presence of the optical lattice does not present technical difficulties
in the compressibility measurements \cite{jin1,jin2},
therefore these measurements \cite{heatcapacity,compres} can be used to detect the uniform compressibility Eq.~\eqref{comp},
the specific heat Eq.~\eqref{spec} and the Wilson ratio in Eq.~\eqref{wilson}.
The various kinds of light and atom scattering methods discussed in \cite{light,annalscaling} is particularly suitable to detect the dynamic compressibility in Eq.~\eqref{scaling}.

\section*{Disscussion}

In this work, we investigate the topological quantum phase transition (TQPT)
of fermions hopping on a honeycomb lattice in the presence of a synthetic non-Abelian gauge potential.
The two Abelian phases $ W=\pm 2 $ are connected by the TQPT tuned by the non-abelian gauge parameters.
We especially distinguish between gauge invariant and gauge dependent quantities across the TQPT.
In fact, the ``Abelian path'' discussed in \cite{Bermudez2010} is just equivalent to the Abelian point $ W=2 $ in Fig.1b.
The positions of the Dirac cones along the ``Abelian path ''  shown in the Fig.~7 in
\cite{Bermudez2010} are gauge dependent quantities and can be shifted by gauge transformations,
but the ground state energy $ \mathcal{E}(\Delta) $ is gauge invariant.
In the TQPT in an anisotropic honeycomb lattice studied in \cite{Tarruell2012,tqcp}, there is no synthetic gauge potential,
the collision is between two time-reversal related Dirac points,
so the merging points can only be located at half of a reciprocal lattice. Here, the collision shown in Fig.2 is {\sl not } between
two time-reversal related Dirac points. The locations of the two merging points $ {\bf P =- Q} $ and the four Dirac points
$ {\bf K}_{1}= -{\bf K}_{3}, {\bf K}_{2}=-{\bf K}_{4} $ are gauge dependent.
But the total number of Dirac points $ N_D $, the colliding process and the TQPT shown in Fig.2 are gauge invariant.
In the 3 dimensional TQPT driven by a Zeeman field discussed in \cite{Volovik2003}, there is no synthetic gauge potential either,
the time-reversal symmetry is broken by the Zeeman field,
the collision is between one left handed and one right-handed Weyl fermions at 3d.
At the BCS mean field level, the critical effective field theory is a 4-component Dirac fermion at 3d  which is different from Eqn.\ref{eq:aHeff}.
So it is a different class of TQPT than that discussed in this paper.
As stressed in this work, in principle, the cold atom experiments \cite{Lin1,Lin2,Lin3,Lin4,Sau2011,Anderson2012,Niu2010,Dalibard2011} can generate various gauges corresponding to the same $ W $ and $ W_{1,2,3} $, so both gauge invariant and gauge dependent quantities can be detected in such experiments.
In sharp contrast, only gauge-invariant quantities can be detected in condensed matter experiments 
(For the discussions on gauge invariant Green functions in high temperature superconductors, see  \cite{gaugeinv1,gaugeinv2,gaugeinv3}).
Indeed, the cold atom experiments of generating synthetic gauge potentials on an optical lattice can
lead to new types of TQPT and also offer new opportunities to explore both gauge invariant and non-gauge invariant quantities through the TQPT.

\section*{Methods}

\noindent \textbf{Derviation of low-energy effective Hamiltonian.}
We first find the energy bands by diagonalizing Hamiltonian matrix [Eq.~\eqref{eq:TBU}]
at the critical point $ \beta_c=\pi/3$
and the merging point ${\bf P}=(\frac{\pi}{2},-\frac{\pi}{2\sqrt{3}})$.
The result is
$S_P^\dagger H(P)S_P=\diag(2t,-2t,0,0)$
and $\Phi_P=S_P^\dagger\Psi_P$,
where
\begin{align}
S_P=
\begin{pmatrix}
i\tfrac{1-\sqrt{3}}{4}      &i\tfrac{-1+\sqrt{3}}{4}    &0&-\tfrac{1+\sqrt{3}}{2\sqrt{2}}\\
-i\tfrac{1+\sqrt{3}}{4}     &i\tfrac{1+\sqrt{3}}{4}     &0&\tfrac{-1+\sqrt{3}}{2\sqrt{2}}\\
\tfrac{1+\sqrt{3}}{4}       &\tfrac{1+\sqrt{3}}{4}      &\tfrac{1-\sqrt{3}}{2\sqrt{2}}&0\\
\tfrac{-1+\sqrt{3}}{4}      &\tfrac{-1+\sqrt{3}}{4}     &\tfrac{1+\sqrt{3}}{2\sqrt{2}}&0\\
\end{pmatrix}.
\end{align}
Then around the $ \beta_c $ and near the $ {\bf P} $,
we can separate the $4\times4$ Hamiltonian into $2\times2$ blocks as
$\tilde{H}(k)=S_P^\dagger H(k)S_P=\left(
\begin{matrix}
H_H &H_C\\
H_C^\dagger &H_L
\end{matrix}\right)$ and
$\Phi(k)=S_P^\dagger\Psi(k)=
(\phi_H ~\phi_L),$
where the upper left diagonal block $H_H$ is the high-energy component,
the lower right diagonal block $H_L$ is the low-energy component,
the off-diagonal blocks $H_C$ is the coupling between the two components
and $\Phi$ is the corresponding field operator.
In the path integral, the quantum partition function is
\begin{align}
Z=Z_0^{-1}\int\mathcal{D}[\bar{\Phi},\Phi]
e^{i\sum_{\omega,k}\bar{\Phi}(\hbar\omega-\tilde{H}(k))\Phi}.
\end{align}
In order to obtain low-energy EFT,
we integrate out high-energy component $\phi_H$
\begin{align}
Z=Z_0^{\prime -1}\int\mathcal{D}[\bar{\phi}_L,\phi_L]
e^{i\sum_{\omega,k}\mathcal{L}_{\rm eff}[\bar{\phi}_L,\phi_L]},
\end{align}
where
$\mathcal{L}_{\rm eff}=\phi_L^\dagger[\hbar\omega-H_L-H_C^\dagger(\hbar\omega-H_H)^{-1}H_C]\phi_L$.
Since $|\omega|\ll t$ and $H_H\propto t$,
we can expand $(\omega-H_H)^{-1}$ in $t^{-1}$ and keep only terms up to $t^{-1}$.
After Legendre transform, we obtain the effective two bands Hamiltonian as
$\mathcal{H}_{\rm eff}
=
\phi_L^\dagger[H_L-H_C^\dagger H_H^{-1}H_C]\phi_L$.
Now we perform an expansion of the Hamiltonian
around the merging point $ {\bf P} $ by writing $ {\bf k} ={\bf P}+ {\bf q}'$ with $|{\bf q}'|\ll1/a$.
Furthermore, we make a $\pi/6$ counter-clockwise rotation 
$
\begin{pmatrix}
q_x\\
q_y\\
\end{pmatrix}
=R_{\pi/6}
\begin{pmatrix}
q'_x\\
q'_y\\
\end{pmatrix}
=
\begin{pmatrix}
\frac{\sqrt{3}}{2} &\frac{1}{2}\\
-\frac{1}{2}       &\frac{\sqrt{3}}{2}\\
\end{pmatrix}
\begin{pmatrix}
q'_x\\
q'_y\\
\end{pmatrix}
$.
After keeping only lowest order derivative terms, we obtain Eq.~\eqref{eq:aHeff}
\begin{align}
\mathcal{H}_{\rm{eff}}({\bf q})
= \phi_L^{\dagger}({\bf q}) [ v\hbar q_y\sigma_x+(\frac{\hbar^2q_x^2}{2m}+\Delta)\sigma_y ]  \phi_L({\bf q}),
\end{align}
where $v=\frac{3ta}{2\hbar}, m=\frac{\hbar^2}{3ta^2}, \Delta=\frac{\sqrt{3}t}{2}(\frac{\pi}{3}-\beta)$.
Relation between effective field
$\phi_L=(\phi_1 ~\phi_2)^T$ and original lattice fields
is give by the unitary matrix $S_P$ as
\begin{align}
\phi_1({\bf q})
=\tfrac{1-\sqrt{3}}{2\sqrt{2}}c_{B}({\bf k}\!\uparrow)+\tfrac{1+\sqrt{3}}{2\sqrt{2}}c_{B}({\bf k}\!\downarrow),~
\phi_2({\bf q})
=-\tfrac{1+\sqrt{3}}{2\sqrt{2}}c_{A}({\bf k}\!\uparrow)+\tfrac{-1+\sqrt{3}}{2\sqrt{2}}c_{A}({\bf k}\!\downarrow),
\label{latticefields}
\end{align}
where  ${\bf q}=R_{\pi/6} {\bf (k-P)} $.

\noindent \textbf{Zero temperature critical exponents.}
The gauge invariant ground state energy density $\mathcal{E}(\Delta)$ of Eq.~\eqref{eq:aHeff} can be written as:
$\mathcal{E}(\Delta)
=\frac{1}{4\pi^2}\int_{-\Lambda_x}^{\Lambda_x} dq_x
 \int_{-\Lambda_y}^{\Lambda_y} dq_y\epsilon_{-}(q_x,q_y;\Delta)$,
where $\Lambda_x$ and $\Lambda_y$
are ultraviolet moment cutoff for $q_x$ and $q_y$ respectively.
To evaluate such a double integral,
we first integrate with respect to $q_y$ variable,
$\int_{-\Lambda_y}^{\Lambda_y}\! dq_y\epsilon_{-}\!
=
f_1(q_x,\Delta)+f_2(q_x,\Delta)+f_3(q_x,\Delta)$,
where we have defined
$f_1(q_x,\Delta)=-\Lambda_y\sqrt{(\frac{\hbar^2q_x^2}{2m}+\Delta)^2+v^2\hbar^2\Lambda_y^2}$,
$f_2(q_x,\Delta)=\frac{1}{2 v\hbar}(\frac{\hbar^2q_x^2}{2m}+\Delta)^2\ln(\frac{\hbar^2q_x^2}{2m}+\Delta)^2$,
and
$f_3(q_x,\Delta)=-\frac{1}{v\hbar}(\frac{\hbar^2q_x^2}{2m}+\Delta)^2
\ln[v\hbar\Lambda_y+\sqrt{(\frac{\hbar^2q_x^2}{2m}+\Delta)^2+v^2\hbar^2\Lambda_y^2}]$.
Due to $\Lambda_y\neq0$ feature, singular behaviors are only hidden in $f_2$.
In the next step, we need to handle the following integration:
$I_2(\Delta)=\int_{-\Lambda_x}^{\Lambda_x} dq_x f_2(q_x,\Delta)$.

Let us take derivative before integration,
since we have the following simple relation
$\frac{\partial^2}{\partial\Delta^2}
f_2(q_x,\Delta)
=
\frac{1}{v\hbar}
\left[
3+\ln(\frac{\hbar^2q_x^2}{2m}+\Delta)^2
\right]$.
For the $\Delta>0$ case, we obtain
\begin{align}
\frac{\partial^2}{\partial\Delta^2}I_2
=
\frac{\sqrt{2m}}{v\hbar^2}
\left[
-2\frac{\hbar\Lambda_x}{\sqrt{2m}}
+4\frac{\hbar\Lambda_x}{\sqrt{2m}}\ln(\frac{\hbar\Lambda_x^2}{2m}+\Delta)
+8\sqrt{\Delta}\arctan\frac{\hbar\Lambda_x}{\sqrt{2m\Delta}}
\right];
\label{eq:app:gt}
\end{align}
for the $\Delta<0$ case, we obtain
\begin{align}
\frac{\partial^2}{\partial\Delta^2}I_2
=
\frac{\sqrt{2m}}{v\hbar^2}
\left[
-2\frac{\hbar\Lambda_x}{\sqrt{2m}}
+4\frac{\hbar\Lambda_x}{\sqrt{2m}}\ln(\frac{\hbar\Lambda_x^2}{2m}+\Delta)
-8\sqrt{-\Delta}\arctanh\frac{\sqrt{-2m\Delta}}{\hbar\Lambda_x}
\right].
\label{eq:app:ls}
\end{align}
Combining Eq.~\eqref{eq:app:gt} and Eq.~\eqref{eq:app:ls},
we have following result for $\partial^2 I_2/\partial\Delta^2$
around the critical point $\Delta=0$,
\begin{align}
\frac{\partial^2}{\partial\Delta^2}I_2
\sim
\begin{cases}
-4\pi\frac{\sqrt{2m}}{v\hbar^2}\sqrt{\Delta}, & {\rm for}~\Delta>0 \\
8\frac{\sqrt{2m}}{v\hbar^2}\frac{\sqrt{2m}\Delta}{\hbar\Lambda_x}. & {\rm for}~\Delta<0
\end{cases}
\end{align}
Notice that $f_1$ and $f_2$ also have linear contributions,
carefully adding these contributions we arrive at the final expression Eq.~\eqref{cusp}
\begin{align}
\mathcal{E}''(\Delta)
\sim
\begin{cases}
-\frac{1}{\pi}\frac{\sqrt{2m}}{v\hbar^2}\sqrt{\Delta}, & {\rm for}~\Delta>0 \\
\frac{\sqrt{2}}{3\pi^2}K(\frac{1}{2})\frac{\sqrt{2m}}{v\hbar^2}\frac{\Delta}{\sqrt{\Lambda}}, & {\rm for}~ \Delta<0
\end{cases}
\end{align}
where $K(z)$ is the complete elliptic integral of the first kind,
$K(1/2)\approx1.85$
and $\Lambda\sim\sqrt{(\frac{\hbar\Lambda_x^2}{2m})^2+v^2\hbar^2\Lambda_y^2}$
is an ultraviolet energy cutoff.

\noindent \textbf{Finite temperature effect.}
From Eq.~\eqref{eq:f} we can directly obtain specific heat as
\begin{equation}
C_v=-T\frac{\partial^2\mathcal{F}}{\partial T^2}
=2\int\frac{d{\bf q}^2}{(2\pi)^2}
\frac{e^{\epsilon_+/(k_BT)}}{(e^{\epsilon_+/(k_BT)}+1)^2}
\frac{\epsilon_+^2}{k_B^2T^2}.
\label{cv}
\end{equation}
From Eq.~\eqref{eq:aHeff}, we can get the fermion Green function
\begin{equation}
G({\bf q},i\omega_n)
=\sum_{s=\pm}\frac{P_s({\bf q})}{i\hbar\omega_n-\epsilon_s({\bf q})},
~
P_s({\bf q})
=\frac{1}{2}
[\sigma_0
+s\frac{v\hbar q_y}{|\epsilon_s({\bf q})|}\sigma_x
+s\frac{\hbar^2 q_x^2/(2m)+\Delta}{|\epsilon_s({\bf q})|}\sigma_y],
\label{green}
\end{equation}
where $P_s({\bf q})$ are the project operators for the $ s=\pm $  band.
The dynamical compressibility can be expressed as
\begin{equation}
\kappa({\bf p},i\omega_n)=-k_BT\sum_{{\bf q},i\nu_n}
\Tr[\mathcal{G}^{(0)}({\bf q}+{\bf p},i\omega_n+i\nu_n)\mathcal{G}^{(0)}({\bf q},i\nu_n)].
\end{equation}
Working out the Matsubara frequency summation and trace, we obtain
\begin{equation}
\kappa({\bf p},i\omega_n)
=\!\sum_{{\bf q},s,s'}M_{ss'}({\bf p,q})
\frac{n_F(\epsilon_{s}({\bf q}))-n_F(\epsilon_{s'}({\bf q+p}))}
{i\omega_n-\epsilon_{s'}({\bf q+p})+\epsilon_s({\bf q})}
\label{kappa},
\end{equation}
where $n_F$ is Fermi distribution function and $M_{ss'}(p,q)$ is
\begin{align}
M_{ss'}(p,q)=&\frac{1}{2}
\left(
1+ss'
\frac{v^2\hbar^2 q_y(q_y+p_y)}
{|\epsilon_{s}(q)\epsilon_{s'}(q+p)|}
+ss'
\frac{(\frac{\hbar^2q_x^2}{2m}+\Delta)[\frac{\hbar^2(q_x+p_x)^2}{2m}+\Delta]}
{|\epsilon_{s}(q)\epsilon_{s'}(q+p)|}
\right).
\end{align}

Explicit evaluations of Eq.~\eqref{cv} and analytical continuations in Eq.~\eqref{green},
\eqref{kappa} lead to the explicit forms of the scaling functions in
 the retarded single particle Green function, the dynamical compressibility and the specific heat in Eq.~\eqref{scaling}.
 The analytic expressions for the scaling function $ \Psi_{i} $ in  $C_{v}$  and $ \Omega_{i} $ in $ \kappa_{u}(T)$ are found to be:
\begin{align}
&\Psi_1(s)=
\frac{4}{\pi^2}\int_{s/2}^\infty dx
\frac{x^{5/2}}{\cosh^2x}
K\left(\sqrt{\tfrac{2x-s}{4x}}\right)  ,\nonumber \\
&\Psi_2(s)=
\frac{4}{\pi^2}\int_{0}^{s/2} dx
\frac{2x^{3}}{\sqrt{2x+s}\cosh^2x}
K\left(\sqrt{\tfrac{4x}{2x+s}}\right)+
\frac{4}{\pi^2}\int_{s/2}^\infty dx
\frac{x^{5/2}}{\cosh^2x}
K\left(\sqrt{\tfrac{2x+s}{4x}}\right)  ,\nonumber \\
&\Omega_1(s)=
\frac{1}{\pi^2}\int_{s/2}^\infty dx
\frac{\sqrt{x}}{\cosh^2x}
K\left(\sqrt{\tfrac{2x-s}{4x}}\right) ,\nonumber \\
&\Omega_2(s)=
\frac{1}{\pi^2}\int_{0}^{s/2} dx
\frac{2x}{\sqrt{2x+s}\cosh^2x}
K\left(\sqrt{\tfrac{4x}{2x+s}}\right)+
\frac{1}{\pi^2}\int_{s/2}^\infty dx
\frac{\sqrt{x}}{\cosh^2x}
K\left(\sqrt{\tfrac{2x+s}{4x}}\right).
\end{align}
where $K(z)$ denote the complete elliptic integrals of the first kind.
Their values in the three regimes shown in the Fig. 4 are listed in Eq.~\eqref{comp}, \eqref{spec}.

\noindent \textbf{Renormalization group analysis of short-range interaction and quenched disorders }
  We can write Eq.~\eqref{eq:aHeff} in the action form:
  \begin{equation}
{\cal S}_{0}[\psi] = \int dq_x dq_y d \omega  \Psi^{\dagger}(q_x,q_y, \omega ) [i \omega + v\hbar q_y\sigma_x
   +  (\frac{\hbar^2q_x^2}{2m}+\Delta)\sigma_y ] \Psi(q_x,q_y, \omega)
   \label{effact}
   \end{equation}
 It is easy to see the canonical dimension in $ (q_x,q_y, \omega ) $ space is $ [  \Psi(q_x,q_y, \omega ) ]=-7/4 $.
   Fourier transforming to $ (x,y, \tau ) $ space leads to $ [  \Psi(x,y,\tau ) ]=3/4 $.

   We can add the short-range interaction $ U $ term to Eq.~\eqref{effact},
$ {\cal S}_{I}[\psi] = U \int dx dy  d \tau [\Psi^{\dagger}(x,y, \tau ) \Psi(x,y, \tau )]^{2}$.
   It is easy to see the canonical dimension of the short range interaction  $ [U]=-1/2 < 0 $,
   so it is irrelevant near the TQPT. This is contrasted with the canonical dimension $ [U]_{D}=-1 < 0 $ of the short range interaction near the
   4 Dirac points $ {\bf K}_{1,2,3,4} $ \cite{coul1,coul2}.

   Now we consider quenched disorders in the gauge parameter $ \beta $.
   As indicated in Eq.~\eqref{eq:aHeff}, at a fixed gauge along the dashed line in Fig.~1(b),
   the tuning parameter $ \Delta \sim \pi/3 - \beta  $, so the randomness in
   the gauge parameter $ \beta $ will lead to the randomness in $ \Delta $. Similarly, the randomness in
   the gauge parameters $ \alpha $ and $ \gamma $ will also lead to random distributions in $ \Delta $ in other gauges.
   We assume all the quenched disorder satisfies a Gaussian distribution with zero mean and variance $ \Delta_{g} $:
   $ \langle g(\vec{r})g(\vec{r}^{\prime}) \rangle_{av}=\Delta_{g}\delta^{2}(\vec{r}-\vec{r}^{\prime}) $
   where $ g $ stands for the gauge parameters $ \alpha, \beta, \gamma $. Averaging over the disorders lead to:
$
   {\cal S}_{g} [\psi]  =  \Delta_{g} \int dx dy d \tau d \tau^{\prime}
 [\Psi^{\dagger}(x,y, \tau ) \sigma_{\alpha} \Psi(x,y, \tau )]^{2}[\Psi^{\dagger}(x,y, \tau^{\prime} ) \sigma_{\alpha} \Psi(x,y, \tau^{\prime} )]^{2}
$.   
 By using the canonical dimension $ [  \Psi(x,y,\tau ) ]=3/4 $,
   one can see the canonical dimension of the short range disorder  $ [\Delta_{g}]=1/2 > 0 $,
   so it is relevant near the TQPT. This is contacted with the canonical dimension  $ [\Delta_{g}]_{D}= 0 $ of the short range disorders near the
   4 Dirac points  \cite{coul1,coul2} $ {\bf K}_{1,2,3,4} $ which is marginal.
   A RG analysis at one loop is needed to determine its fate \cite{coul1,coul2}.



\begin{addendum}

\item [Acknowledgement]

We acknowledge helpful discussions with
Y.-X. Yu, S.-J. Jiang, H. Pu, Carlos A. R. S\'{a} de Melo,
Alexei M. Tsvelik and S. P. Kou.
This work was supported by the NKBRSFC under Grants
No. 2010CB922904, No. 2011CB921502, and No. 2012CB821300,
NSFC under Grants No. 10934010.
J. Ye was supported by NSF-DMR-1161497, NSFC-11074173, -11174210,
Beijing Municipal Commission of Education under Grant No. PHR201107121.

\item [Author Contributions]
All authors planned and designed theoretical numerical studies.
All contributed in completing the paper.

\item [Competing Interests]
The authors declare that they have no competing financial interests.

\item [Correspondence]
Correspondence and requests for materials should be addressed to Sun, Fadi.
\end{addendum}

\clearpage

\newpage
\bigskip
\textbf{Figure 1 Lattice geometry and phase diagram.}
(a) The honeycomb lattice consists of sublattice
$ A $ (red dots) and sublattice $ B $ (blue dots).
The up and down arrows represent the spin degrees of freedom. $a$ is the lattice constant.
The non-Abelian gauge potentials $ U_{1,2,3} $ with directions are displayed on the three links inside the unit cell.
(b) The phase diagram of our system as a function of gauge parameters $\alpha$ and $\beta$.
The yellow (green) region has $ N_D=8 $ ($ N_D=4 $)  Dirac points shown in the insets.
The center $C$ point is the $ \pi $ flux  Abelian point.
The 4 edges of the square belong to the gauge equivalent trivial Abelian point.
We investigate the topological quantum phase transition from $C$ point to $D$ point along the dashed line.

\bigskip

\textbf{Figure 2 Topologies of different Fermi surface.}
The different Fermi surface topologies of the $\epsilon_{1-}$ in the 1st Brillouin zone along the dashed line in the Fig.~1(b).
(a) The $ \pi $ flux Abelian point $\alpha=\pi/2$, $\beta=\pi/2$ inside the $ N_D=8 $ phase,
(b) The $\alpha=\pi/2$, $\beta=2\pi/5$ inside the $ N_D=8 $ phase,
(c) The TQPT at $\alpha=\pi/2$, $\beta_c=\pi/3$. The two emerging points are located
at $ {\bf P}= (\frac{\pi}{2},-\frac{\pi}{2\sqrt{3}}) $ and its time-reversal partner  $ {\bf Q}=-{\bf P}$.
The four Dirac points are located at
${\bf K}_1 =(\frac{5\pi}{12},\frac{\pi}{4\sqrt{3}})=-{\bf K}_3,
{\bf K}_2 =(-\frac{\pi}{12},\frac{\sqrt{3}\pi}{4})=-{\bf K}_4 $.
(d) The $\alpha=\pi/2$, $\beta=\pi/4$ inside the $ N_D=4 $ phase.

\bigskip

\textbf{Figure 3 Ground-state energy density.}
(a)The ground-state energy density on the lattice scale $\mathcal{E}_{\rm latt}(\beta)$ as a function of $\beta$.
(b)The first-order derivative of the ground-state energy density on the lattice scale $\mathcal{E}_{\rm latt}(\beta)$ with respect to $\beta$.
(c) The second-order derivative of the ground-state energy density on the lattice scale $\mathcal{E}_{\rm latt}(\beta)$ with respect to $\beta$.
It shows a cusp when $\beta=\pi/3,2\pi/3$.
(d) The third-order derivative of $\mathcal{E}_{\rm latt}(\beta)$ with respect to $\beta$.
It shows discontinuity when $\beta=\pi/3,2\pi/3$,
so the system undergoes a third order topological quantum phase transition.

\bigskip

\textbf{Figure 4 Finite-$T$ Phase diagram.}
(a) The gauge-invariant phase diagram in terms of the Wilson loops $ W $ and $ W_1 $. 
The yellow (green) regime is $ N_D=8 $ ($ N_D=4 $).
The dashed line corresponds to the one in Fig. 1(b).
(b) Finite-$T$ Phase diagram of the topological quantum phase transition as a function of the flux $\Delta$ and the temperature $T$.
There is a topological quantum phase transition at $ T=0, \Delta=0 $. The two dashed lines stand for the crossovers at $ T \sim |\Delta| $.

\newpage
\begin{figure}
\begin{center}
\epsfig{file=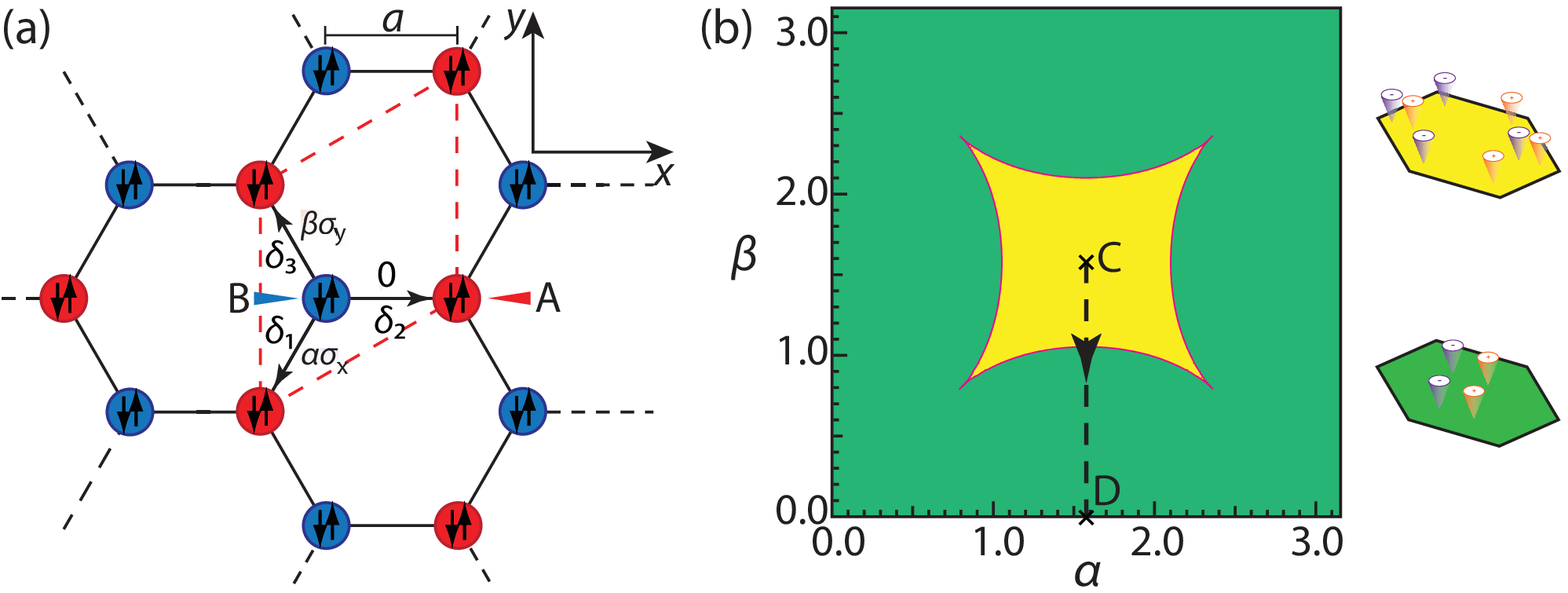,width=14cm}
\end{center}
\label{fig:HLPD}
\end{figure}

\begin{figure}
\begin{center}
\epsfig{file=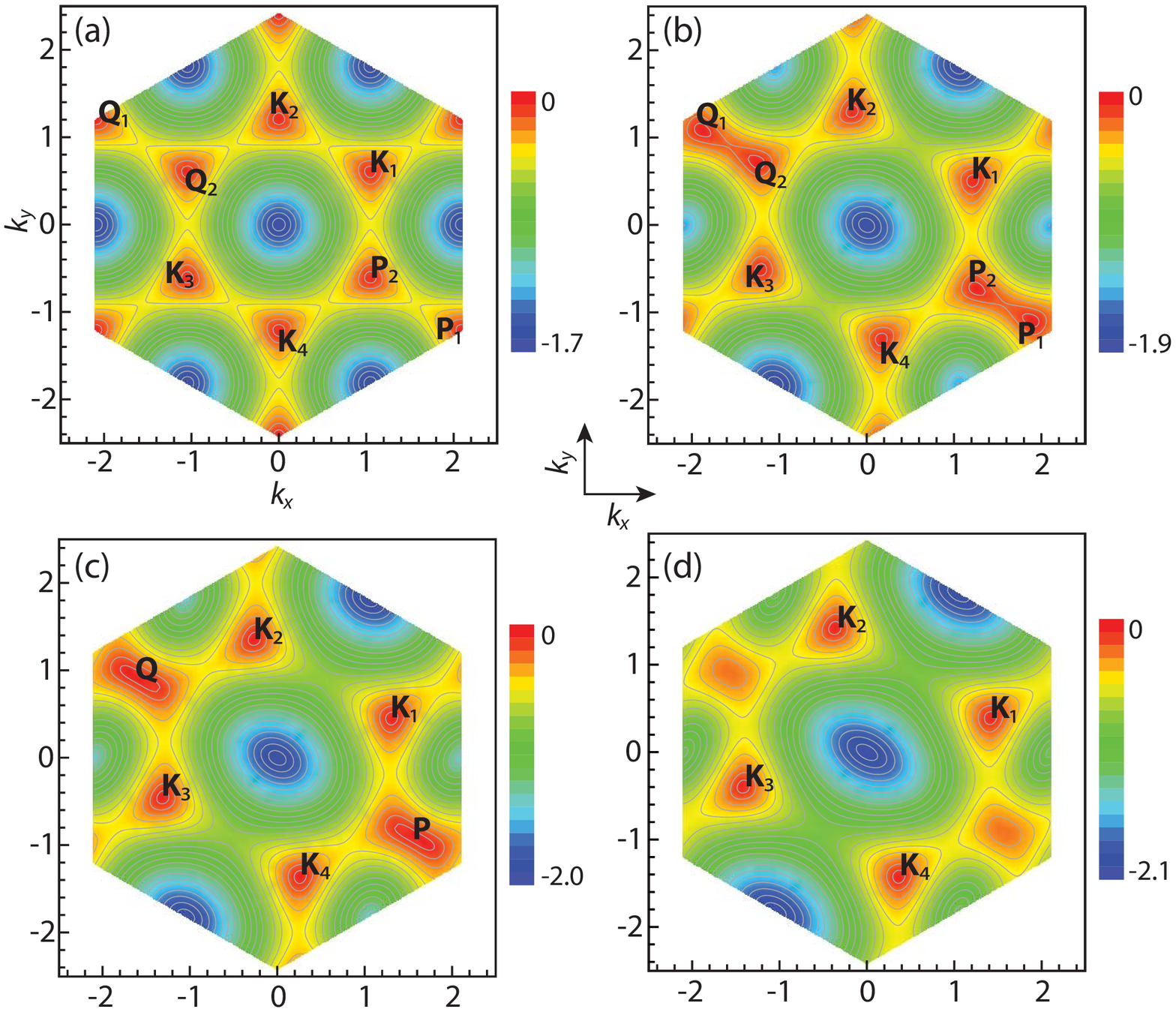,width=14cm}
\end{center}
\label{fig:ES}
\end{figure}

\begin{figure}
\begin{center}
\epsfig{file=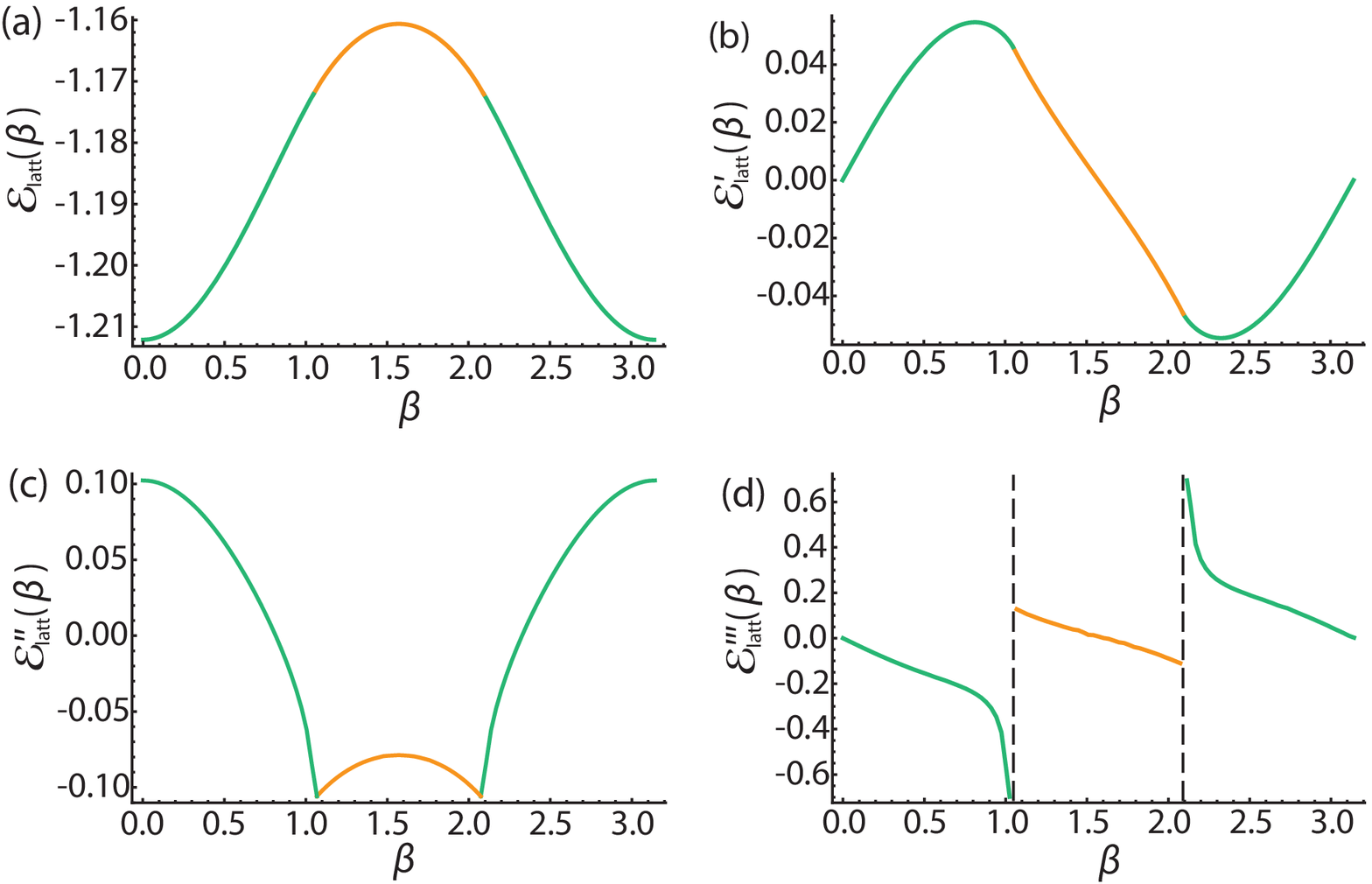,width=15cm}
\end{center}
\label{fig:TQPT}
\end{figure}

\begin{figure}
\begin{center}
\epsfig{file=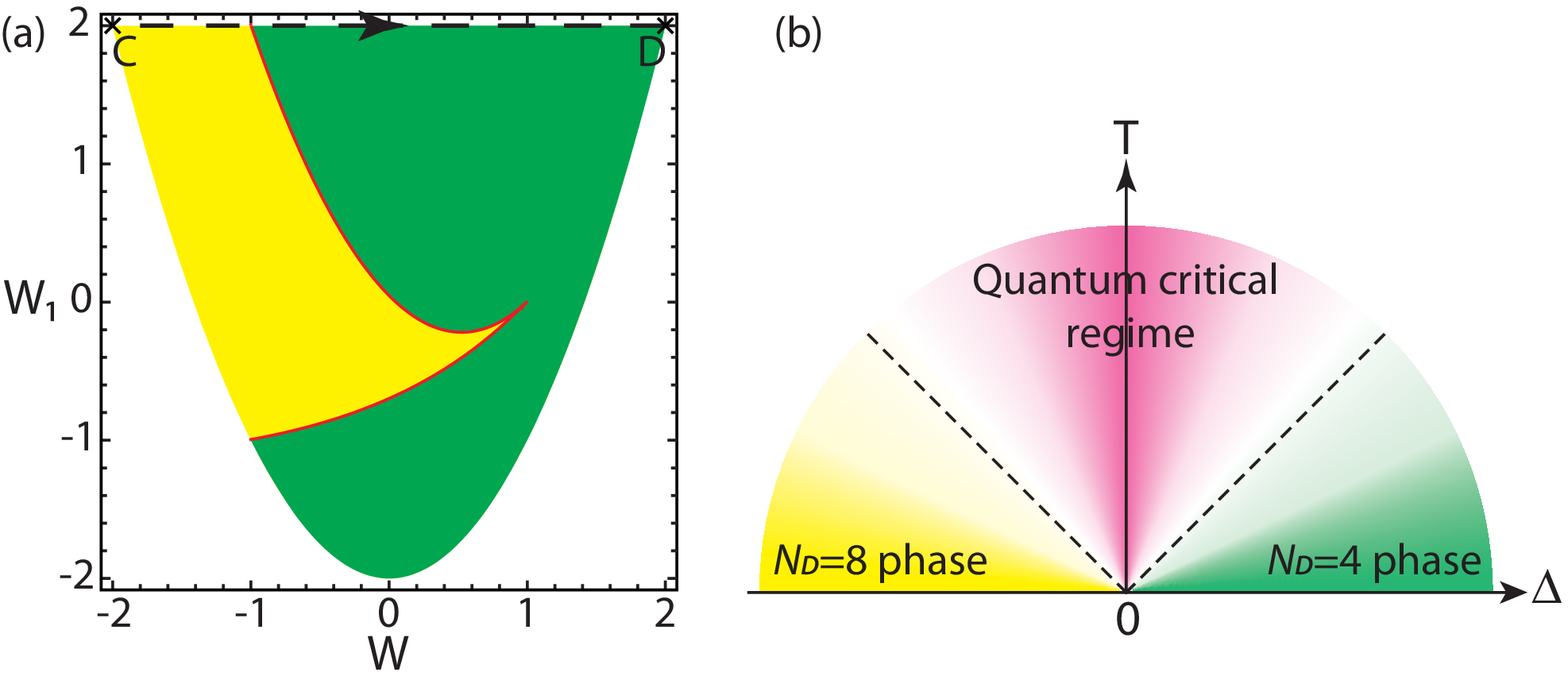,width=15cm}
\end{center}
\label{fig:FiniteT}
\end{figure}
\clearpage
\bigskip

\end{document}